\documentclass[a4paper,10pt]{article}
\usepackage[dvips]{graphicx}
\usepackage{amssymb,amsmath}
\numberwithin{equation}{section}

\begin{document}

\title{ Some Models of Cyclic and  Knot Universes }

\author{K.R. Yesmakhanova,   N. A. Myrzakulov, K. K. Yerzhanov, G.N. Nugmanova, \\N.S. Serikbayaev and  R. Myrzakulov\footnote{The corresponding author. Email: rmyrzakulov@csufresno.edu; rmyrzakulov@gmail.com }   \\ \textit{Eurasian International Center for Theoretical Physics,  } \\ \textit{Eurasian National University, Astana 010008, Kazakhstan} } 

\date{}

\maketitle
\begin{abstract}In this paper, we obtained a class of
oscillatory, cyclic and knot type solutions from the non-linear
Friedmann equations. This is performed by choosing specific forms of
energy density and pressure of matter. All the expressions written
here are in dimensionless form. We show that evolutionary path taken
by the spatial coordinates in the model follow various knots,
specifically trefoil and eight-knots. We provide several examples
and plot relevant cosmological parameters in figures. Our cyclic
models can be interpreted as a periodic cosmological model, such
that early and late time acceleration are unified under the same
mechanism. Finally we have presented some examples of knot universes for the Bianchi - I spacetime.
\end{abstract}
\tableofcontents
\newpage
\section{Introduction}

The understanding of formation and evolution of the observable
universe is the main theme of modern cosmology. About the former
problem, there are several candidates including the standard
cosmological model which predicts a singularity out of which our
universe was born, while it predicts nothing about what happened
before that. Other models like braneworlds and cyclic models predict
a cyclic nature of cosmic evolution, according to which universe
evolves from one singular state (big bang) to another one
\cite{cyclic}. Another interesting theoretical problem is the
phenomenon of cosmic acceleration occurring at present time, usually
attributed to `dark energy' (DE). There are numerous approaches to
study this phenomena \cite{sami}, however, it is fair to say that
none of them is particularly well motivated, and many of them appear
like acts of desperation on the part of theorists. From this
situation we sometimes come to the conclusion that perhaps we must
return to the origin and study in detail the mathematical nature  of
the basic gravitational equations. Following this idea we have
studied some integrable and non-integrable reductions of the
Einstein equation (Friedmann equation) in \cite{MR1}-\cite{MR3}. In
\cite{Kuralay}, we have established the relationship between
solutions of the Einstein equation and the Ramanujan and  Chazy
equations.

Returning to a cyclic scenario which in fact was proposed long time ago \cite{Tolman}   we note that the idea to consider the whole universe as one cyclic system has already attracted many authors. In particular, it has achieved a lot of successes in recent decades (see e.g. Refs. \cite{Cai1}-\cite{Piao} and references therein). One of interesting aspects of cyclic scenario is its possible connection with the knot scenario   proposed in  \cite{Rat} (see also Ref. \cite{Kuralay}).  One of  important branches of mathematics (algebraic geometry), the knot theory  describe a debth properties of the 3-dimensional space \cite{tait}-\cite{knot}. So that these arguments in our opinion give us physical and mathematical motivations in order to futher investigation the relationship between cyclic universe models and  knot universe models. 
In this paper, we study in more detail the
 relationship between some models of the cyclic universe and the knot theory. 
As examples we consider the trefoil  and   figure-eight knot universe models.

The paper is organized as follows. In Sec. II we briefly give some
 basic equations of the Einstein gravity. Sec. III is devoted to
 study the relation between the Friedmann equation with some Equation
  of state and the trefoil knot and Sec. IV with the figure-eight knot.
   Some similar  models of
     a cyclic universe  are studied in Sec.V. A new realizations  of knot universes for the Bianchi - I   models   are considered in Sec. VI. The last section is
     devoted to the conclusion.

\section{The model and basic equations}
In this section we briefly review some basic facts about the
Einstein's field equation. We start from the standard gravitational
action (chosen units are $c=8\pi G=1$)
\begin{equation}
S=\frac{1}{4}\int d^{4}x\sqrt{-g}(R-2\Lambda+L_m),
\end{equation}
where $R$ is the Ricci scalar, $\Lambda$ is the cosmological
constant and $L_m$ is the matter Lagrangian. For a general metric
$g_{\mu\nu}$, the line element is
 \begin{equation}
ds^2=g_{\mu\nu}dx^\mu dx^\nu, \ \ \mu,\nu=0,1,2,3
\end{equation}
the corresponding Einstein field equations are given by
 \begin{equation}
R_{\mu\nu}+\Big(\Lambda-\frac{1}{2}R\Big)g_{\mu\nu}=- T_{\mu\nu},
\end{equation}
Here $R_{\mu\nu}$ is the Ricci tensor. This equation forms the
mathematical basis of the theory of general relativity. In (3),
$T_{\mu\nu}$ is the energy-momentum tensor of the matter field
defined as
  \begin{equation}
T_{\mu\nu}=\frac{2}{\sqrt{-g}}\frac{\delta L_m}{\delta g^{\mu\nu}},
\end{equation}
and satisfies the conservation equation
 \begin{equation}
\nabla_\mu T^{\mu\nu}=0,
\end{equation}
where $\nabla_\mu$ is the covariant derivative which is the relevant
operator to smooth a tensor on a differentiable manifold. Eq. (5)
yields the conservations of energy and momentums, corresponding to
the independent variables involved. The general Einstein equation
(3) is a set of non-linear partial differential equations. We
consider the Friedmann-Robertson-Walker (FRW) metric which
represents a spatial and homogeneous spacetime relevant to
cosmological perspective. This metric has the form (in suitable
dimensionless coordinates)
\begin{equation}
ds^2=d\tau^2-a(\tau)^2[(dx'_{1})^{2}+(dx'_{2})^{2}+(dx'_{3})^{2}).
\end{equation}
Here we employed the following notations $x^{'}_1=x_1/x_{10}$,
$x^{'}_2=x_2/x_{20}$, $x^{'}_3=x_{3}/x_{30}$ and $\tau=t/t_0$ for
making the above equation dimensionless. Quantities with subscript 0
refer to their values at present epoch, that is $x_{10}, x_{20},
x_{30}, t_0$ are some constants. We write the corresponding Einstein
equations (Friedmann equation and the conservation equation) in the
following dimensionless form
\begin{eqnarray}
3H^2-\rho&=&0,\\
2\dot{H}+3H^2+p&=&0,\\
\dot{\rho}+3H(\rho+p)&=&0,
\end{eqnarray}
 where $H=(\ln a)_{,\tau}=a_{,\tau}/a$, is the dimensionless Hubble
 parameter and $\dot{H}=dH/d\tau$, $\rho$ is the density of energy and $p$ is the
 pressure. With the use
of dimensionless variables, all the equations in this paper become
dimensionless. Out of the three equations (7)-(9), only two are
independent i.e. (7) and (8) or (7) and (9).

\section{Knot universe: the trefoil knot case}
The aim of this and next sections is to establish a connection
between the cyclic universe models with the knot theory. In this
section we want to construct the simplest examples of the knot
universe, namely, the trefoil knot universe. Let us consider some
explicit examples.

\subsection{Example 1} Let the equation of state (EoS) that is the  pressure $p$ and energy density $\rho$ have the
following forms \cite{Kuralay}
\begin{eqnarray}
\rho(\tau)&=&3r^2\cos^2 (2\tau), \\
p(\tau)&=&6\sin (3\tau)\cos (2\tau)+4r\sin (2\tau)-3r^2\cos^2
(2\tau),
\end{eqnarray}
where
 $$r\equiv 2+\cos (3\tau),$$
  is another dimensionless parameter.
Now substituting these expressions for the energy density and the
pressure into the equations (7)-(8), we get
\begin{equation}
H(\tau)=r\cos (2\tau).
\end{equation}
Such periodic Hubble's parameters are familiar from previous works
on occurrence of an oscillating universe, it has been shown using an
inhomogeneous equation of state for dark energy fluid, the Hubble
parameter  presents a periodic behavior such that early and late
time acceleration are unified under the same mechanism \cite{diego}.

\begin{figure}
\centering
 \includegraphics[scale=0.5] {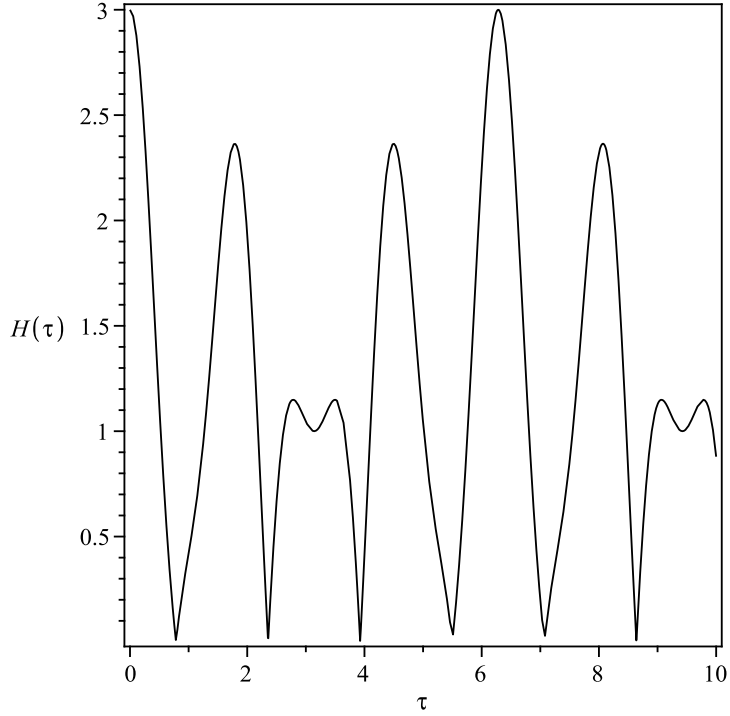}
  \caption{ The Hubble parameter is plotted as given in (12).}
 \label{1.eps}
\end{figure}

Upon integration, we obtain the scale factor for this model
\begin{equation}
a(\tau)=a_0e^{[{0.5\sin (\tau)+\sin (2\tau)+0.1\sin
(5\tau)}]},
\end{equation}
where $a_0=a(\tau=0)$ is the value of scale factor taken at present
time. Having obtained relevant parameters, we can discuss some
cosmological implications of this model. The first quantity is
deceleration parameter $q$ defined as
\begin{equation}
q=-1-\frac{\dot{H}}{H^2}.
\end{equation}
A universe expanding with acceleration (deceleration) has $q<0$
($q>0$). Since the observable universe is currently in accelerated
expansion, we demand $q> -1$. Thus using (12) and (14) we have
 \begin{equation}
  q= -1+{\frac {3\,\sin ( 3\,\tau) \cos ( 2\,\tau ) +2\,
 ( 2+\cos( 3\,\tau )  ) \sin ( 2\,\tau ) }
{ ( 2+\cos ( 3\,\tau )  ) ^{2}  \cos^{2} ( 2 \,\tau ) }} ,
\end{equation}

The EoS parameter for this example is given by
\begin{equation}
\omega\equiv\frac{p}{\rho}=\frac{6\sin (3\tau)\cos( 2\tau)+4r\sin
(2\tau)-3r^2\cos^2 (2\tau)}{3r^2\cos^2 (3\tau)}.
\end{equation}
Figure (1) shows time evolution of the $H$ given by (12). It may be
non differentiable \cite{diego}. It can be used for a periodic
behavior such that early and late time acceleration are unified
under the same mechanism.

Let us introduce three new variables as
\begin{eqnarray}
x&=&H, \\
y&=&\sqrt{r^2-H^2}, \\
z&=& \sqrt{1-(r-2)^2}.
\end{eqnarray}
Using the above expressions of $H$ and $r$, we can rewrite
 equations (17)-(19) as \cite{Kuralay}
\begin{eqnarray}
x&=&r\cos (2\tau), \\
y&=&r\sin (2\tau), \\
z&=&\sin (3\tau).
    \end{eqnarray}
Eqs. (20-22) are nothing but the parametric equations for  the
trefoil knot (see Figure 2). This curve lies entirely on the torus
\begin{equation}
(r-2)^2+z^2=1,
\end{equation}
making trefoil the simplest example of a torus knot. In fact, the
trefoil is a (2,3)-torus knot because the curve winds around the
torus thrice in one direction and twice in the other direction.

\begin{figure}
\centering
 \includegraphics[scale=1.4] {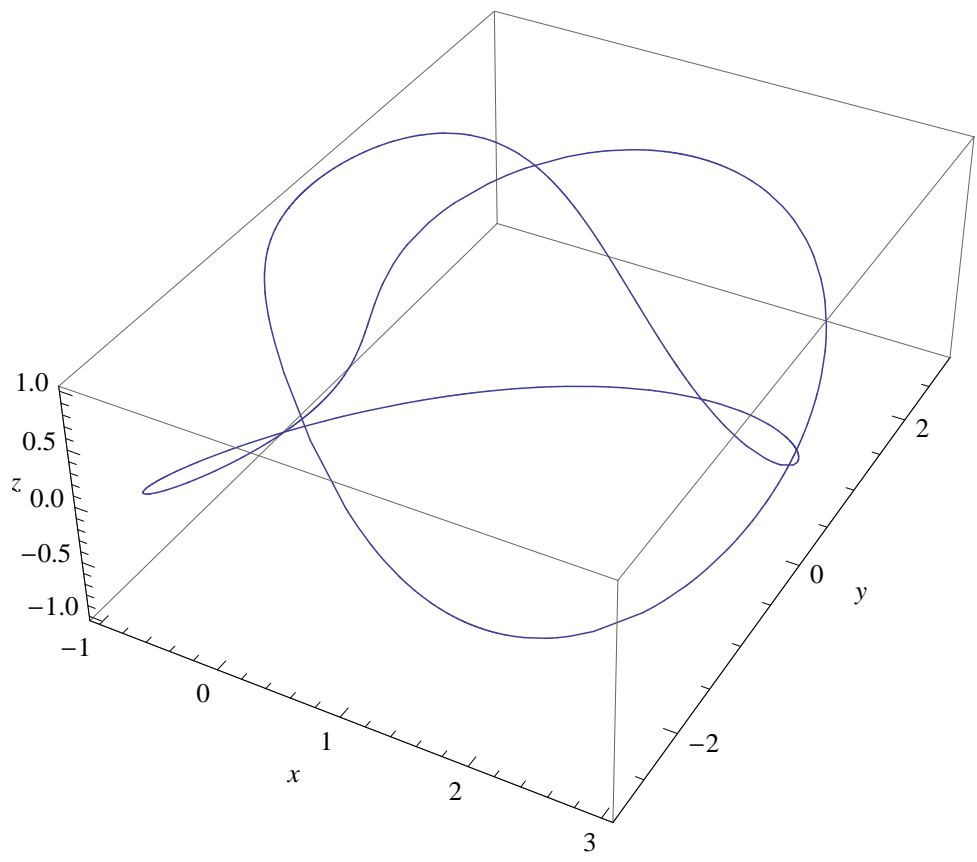}
  \caption{ The trefoil knot for the equations (21)-(23).}
 \label{2.eps}
\end{figure}

\subsection{Example 2} We consider the scale factor of the form
    \begin{equation}
a^{'}=\frac{a}{a_0}=r\cos (2\tau),
\end{equation}
where $a_0=a(t=0)$ and assume $a_0=1$. Thus the Hubble parameter
becomes
\begin{equation}
H=[\ln (r\cos (2\tau))]_{\tau}=-\frac{3\sin (3\tau)\cos(
2\tau)+2(2+\cos (3\tau))\sin (2\tau)}{(2+\cos (3\tau))\cos (2\tau)}.
\end{equation}
In figure 3, we show the time evolution of $\log H$ vs $\tau$. It
describes a periodic behavior such that early and late time
acceleration are unified under the same mechanism.

\begin{figure}
\centering
 \includegraphics[scale=0.5]{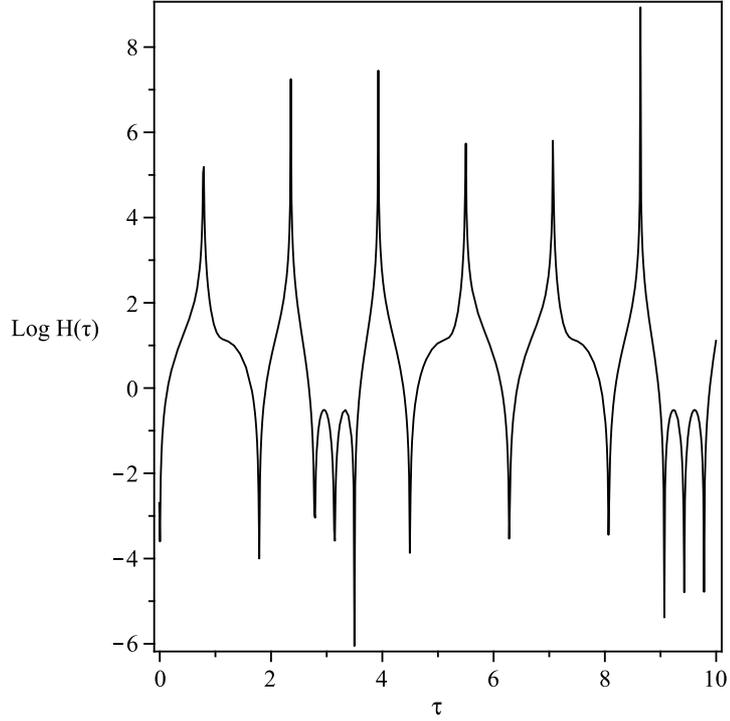}
    \caption{The Log-Hubble parameter given by (25).}
    \label{3.eps}
\end{figure}

Making use of (7), (8) and (25), the energy density and pressure
become
\begin{eqnarray}
    \rho&=&3\{[\ln (r\cos (2\tau))]_{,\tau}\}^2, \\
        p&=&-2[\ln (r\cos (2\tau))]_{,\tau\tau}-3\{[\ln (r\cos (2\tau))]_{,\tau}\}^2.
    \end{eqnarray}
    The corresponding parameter of the EoS is
        \begin{equation}
\omega=-1-\frac{2}{3}\frac{[\ln (r\cos (2\tau))]_{,\tau\tau}}{\{[\ln
(r\cos( 2\tau))]_{,\tau}\}^2}.
\end{equation}
To describe the trefoil knot, we introduce three new variables $x,
y, z$ as:
\begin{eqnarray}
    x&=&a^{'}, \\
        y&=&\sqrt{r^2-a^{'2}}, \\
            z&=& \sqrt{1-(r-2)^2},
    \end{eqnarray}
  where   $a'=a/a_0,\ \ a_0=$constant.
In terms of $\tau$ these functions take the same form as in
(20)-(22) so that they describe the trefoil knot curve. This means
that our model with the EoS (28) corresponds to a knot universe if
be exactly to the trefoil knot universe.

\section{Knot universe: the figure-eight knot case}

Now consider a more  complex case than the previous trefoil knot
universe, namely, the figure-eight knot. Here some examples of such
figure-eight knot universes.

\subsection{Example 1} Let the  EoS is given in the parametric form \cite{Kuralay}
    \begin{eqnarray}
    \rho&=&3h^2\cos^2 (3\tau), \\
        p&=&4\sin (2\tau)\cos (3\tau)+6h\sin (3\tau)-3h^2\cos^2 (3\tau),
    \end{eqnarray}
    where $$h=2+\cos (2\tau).$$ Equations (7)-(8) then give
    \begin{equation}
H=h\cos (3\tau).
\end{equation}
Figure 4 shows the time evolution of the $H$ for this example. It
may be interpreted as a periodic cosmological model, such that early
and late time acceleration are unified under the same mechanism.

\begin{figure}
\centering
 \includegraphics[scale=0.5]{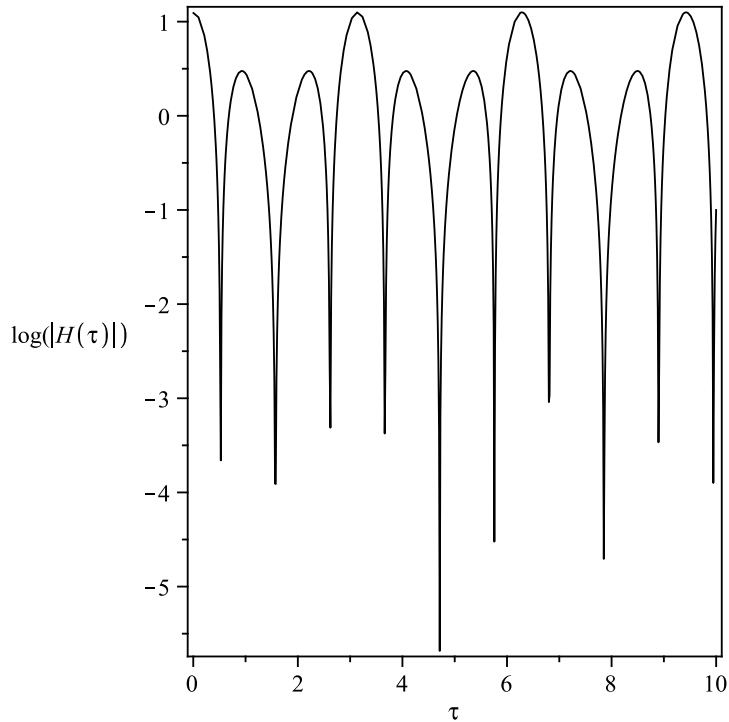}
    \caption{The Log-Hubble parameter given by (34).}
    \label{4.eps}
\end{figure}
This result (34) tells us that the scale factor has the form
    \begin{equation}
a(\tau)=a_0e^{[{0.5\sin (\tau)+\frac{2}{3}\sin (3\tau)+0.1\sin
(5\tau)}]}.
\end{equation}
 The corresponding EoS parameter for this model is
    \begin{equation}
\omega=\frac{p}{\rho}=\frac{4\sin (2\tau)\cos (3\tau)+6h\sin(
3\tau)-3h^2\cos^2 (3\tau)}{3h^2\cos^2( 3\tau)}.
\end{equation}
As above, we introduce three new functions as:
\begin{eqnarray}
    x&=&H, \\
        y&=&\sqrt{h^2-H^2}, \\
            z&=& 2(h-2)\sqrt{1-(h-2)^2}.
    \end{eqnarray}
    This system $x$, $y$, $z$ can be rewritten in parametric form as \cite{Kuralay}
    \begin{eqnarray}
    x&=&h\cos (3\tau), \\
        y&=&h\sin (3\tau), \\
            z&=&\sin (4\tau),
    \end{eqnarray}
which is nothing but the parametric equations of    the figure-eight
knot (see Figure 5). Note that this figure-eight knot satisfies the
equation
\begin{equation}
4(h-2)^4-4(h-2)^2+z^2=0.
\end{equation}
 \begin{figure}
\centering
 \includegraphics[scale=1.2]{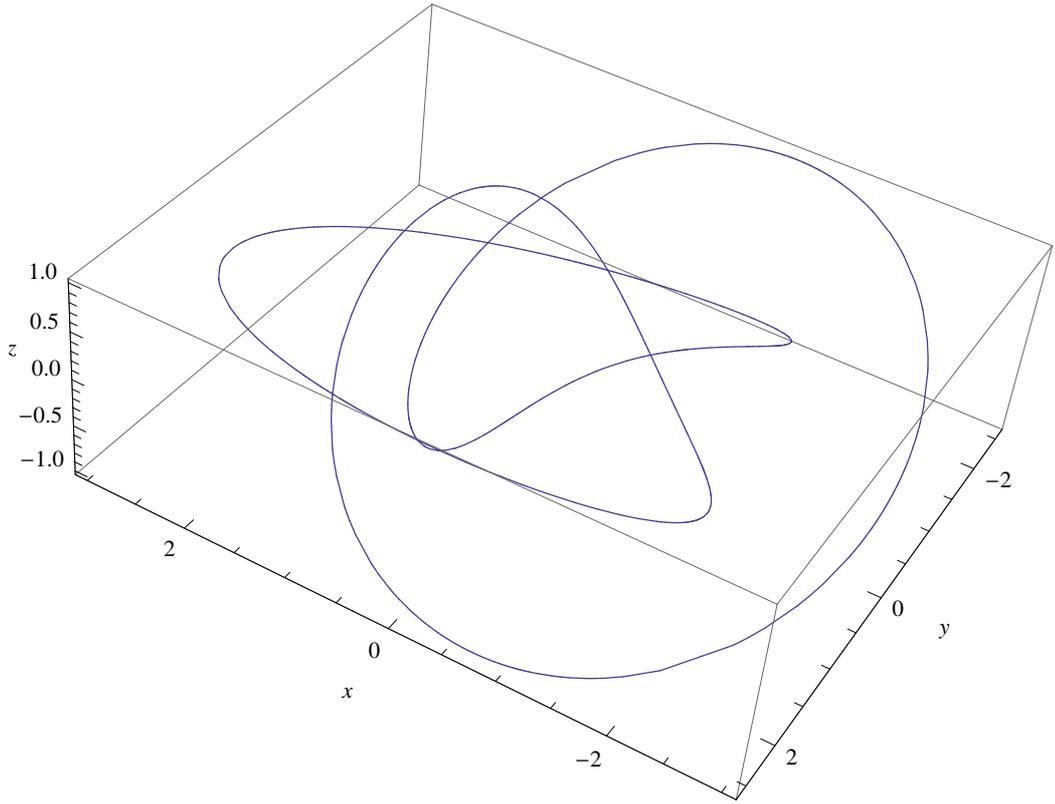}
    \caption{The figure-eight knot for the equations (40)-(42).}
    \label{5.eps}
\end{figure}

\subsection{Example 2} Consider the scale factor of the form
\begin{equation}
a'=h\cos (3\tau)=(2+\cos ((2\tau))\cos (3\tau),
\end{equation}
so that  the Hubble parameter takes the form
\begin{equation}
H=[\ln (h\cos( 3\tau))]_{,\tau}=-\frac{2\sin (2\tau)\cos
(3\tau)-3(2+\cos (2\tau))\sin (3\tau)}{(2+\cos (2\tau))^2\cos^2
(3\tau)}.
\end{equation}
Figure 6 shows the time evolution of the $H$ for this example. It is
a periodic cosmological model, such that the early and late time
accelerations are unified under the same mechanism.

\begin{figure}
    \centering
\includegraphics[scale=0.5]{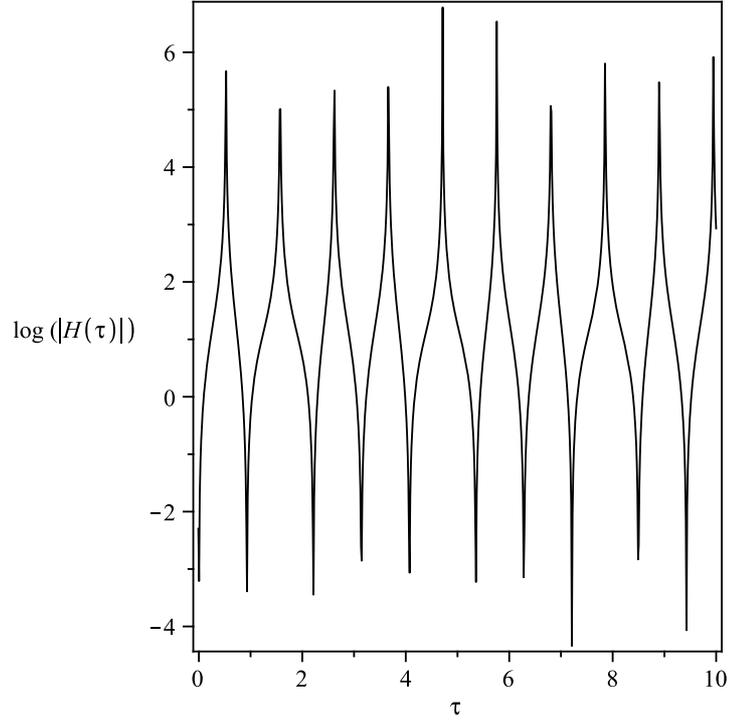}
    \caption{The Hubble parameter given by (46).}
    \label{6.eps}
\end{figure}

Then the  equations (7)-(8) give the following parametric EoS
\begin{eqnarray}
    \rho&=&3\{[\ln (h\cos (3\tau))]_{,\tau}\}^2, \\
        p&=&-2[\ln (h\cos (3\tau))]_{,\tau\tau}-3\{[\ln (h\cos (3\tau))]_{,\tau}\}^2
    \end{eqnarray}
which gives the  parameter of the EoS in the form
        \begin{equation}
\omega=-1-\frac{2}{3}\frac{[\ln (r\cos (2\tau))]_{,\tau\tau}}{\{[\ln
(r\cos (2\tau))]_{,\tau}\}^2}.
\end{equation}
Again  we introduce three new variables $x, y, z$ as:
\begin{eqnarray}
x&=&a', \\
y&=&\sqrt{h^2-a^{'2}}, \\
z&=& 2(h-2)\sqrt{1-(h-2)^2},
\end{eqnarray}
where $a^{'}=a/a_0, \quad a_0=const.$  If we rewrite these equations
in terms of $\tau$ then we get again the system  (40)-(42) which are
the parametric equations for  the figure-eight  knot. So this model
with the EoS (46)-(47) describes  the figure-eight knot universe.

\section{Similar  models of a cyclic universe}

\subsection{Example 1} Let us consider an oscillating universe
described by the Hubble parameter
\begin{equation}
H(\tau)=\alpha+\beta \cos (\nu \tau).
\end{equation}
Then, by substituting this expression into the equations (7), (8),
we obtain
\begin{eqnarray}
\rho&=&3(\alpha+\beta \cos (\nu \tau))^2, \\
p&=&2\beta\nu\sin (\nu \tau)-3(\alpha+\beta \cos (\nu\tau))^2,
    \end{eqnarray}
    so that the EoS parameter becomes
\begin{equation}
\omega=-1+\frac{2\beta\nu\sin (\nu \tau)}{3(\alpha+\beta \cos (\nu
\tau))^2}.
\end{equation}
The corresponding scale factor is
\begin{equation}
a(\tau)=a_0e^{[{\alpha \tau+\frac{\beta}{\nu}\sin (\nu
\tau)}]}.
\end{equation}

\subsection{Example 2} One can consider now an  oscillating universe
described by the Hubble parameter
\begin{equation}
H(\tau)=\beta \sin (\nu \tau),
\end{equation}
where $\beta$ and $\nu$ are constants. Then, by substituting this
expression into the equations (7),(8), we get
\begin{eqnarray}
\rho&=&3\beta^2 \sin^2 (\nu \tau), \\
p&=&-2\beta\nu\cos (\nu \tau)-3\beta^2 \sin^2 (\nu \tau).
\end{eqnarray}
so that the EoS parameter is given by
\begin{equation}
\omega=-1-\frac{2\beta\nu\cos (\nu \tau)}{3\beta^2 \sin^2 (\nu
\tau)}.
\end{equation}
For the scale factor we get
\begin{equation}
a(\tau)=a_0e^{[{-\frac{\beta}{\nu}\cos( \nu \tau)}]}.
\end{equation}

\subsection{Example 3} Let's take
        \begin{equation}
a(\tau)=\alpha+\beta \cos^2 (\nu \tau).
\end{equation}
Then, the Hubble parameter becomes
    \begin{equation}
H(\tau)=-\frac{2\beta\nu\cos (\nu \tau)\sin (\nu \tau)}{\alpha+\beta
\cos^2 (\nu \tau)}.
\end{equation}
 By substituting this expression into the equations (7)-(8), it yields,
\begin{eqnarray}
\rho&=&12\frac{\beta^2\nu^2\cos^2 (\nu \tau)\sin^2 (\nu \tau)}{(\alpha+\beta \cos^2 (\nu \tau))^2}, \\
p&=&-12\frac{\beta^2\nu^2\cos^2 (\nu \tau)\sin^2 (\nu
\tau)}{(\alpha+\beta \cos^2 (\nu
\tau))^2}+\frac{4\beta\nu^2[\beta\cos^4(\nu \tau)+\beta\cos^2(\nu
\tau)\sin^2(\nu \tau)+\alpha(\cos^2(\nu \tau)-\sin^2(\nu
\tau))]}{(\alpha+\beta \cos^2 (\nu \tau))^2}.\nonumber\\
\end{eqnarray}
So that the EoS parameter is given by
            \begin{equation}
\omega=-1+\frac{\beta\cos^4(\nu \tau)+\beta\cos^2(\nu
\tau)\sin^2(\nu \tau)+\alpha(\cos^2(\nu \tau)-\sin^2(\nu
\tau))}{3\beta\cos^2 (\nu \tau)\sin^2 (\nu \tau)}.
\end{equation}

\subsection{Example 4} Our next example is:
        \begin{equation}
a(\tau)=\alpha+\beta \cos (\nu \tau).
\end{equation}
Then, the Hubble parameter becomes
    \begin{equation}
H(\tau)=-\frac{\beta\nu\sin (\nu \tau)}{\alpha+\beta \cos (\nu
\tau)}.
\end{equation}
By substituting this expression into the equations (7)-(8), we
obtain
\begin{eqnarray}
\rho&=&\frac{3\beta^2\nu^2\sin^2 (\nu \tau)}{(\alpha+\beta \cos (\nu \tau))^2}, \\
p&=&-\frac{3\beta^2\nu^2\sin^2 (\nu \tau)}{(\alpha+\beta \cos^2 (\nu
\tau))^2}+\frac{2\beta\nu^2(\beta+\alpha\cos^2(\nu
\tau))}{(\alpha+\beta \cos (\nu \tau))^2}.
\end{eqnarray}
So that the EoS parameter simplifies to
\begin{equation}
\omega=-1+\frac{2}{3\beta}\frac{\beta+\alpha\cos(\nu
\tau)}{\sin^2(\nu \tau)}.
\end{equation}

\subsection{Example 5} General model of a cyclic universe: We introduce
\begin{eqnarray}
\rho&=&\varrho _{{0}}\cos^2 \left( 2\,\pi \,\nu\,\tau+\theta \right)
\\
p&=&-\frac{\sqrt {\varrho _{{0}}}}{3}\, \left( 3\, \left( \cos
\left( 2\,\pi \, \nu\,\tau+\theta \right)  \right) ^{2}\sqrt
{\varrho _{{0}}}-4\pi \,\nu\,\sqrt {3} \sin \left( 2\,\pi
\,\nu\,\tau+\theta \right)  \right).
\end{eqnarray}
We get
\begin{eqnarray}
H=\sqrt{\frac{\varrho _{{0}}\cos \left( 2\,\pi \,\nu\,\tau+\theta
\right)}{3}}.
\end{eqnarray}
Hence we have
\begin{eqnarray}
a(\tau)=a_0e^{[{\sqrt{\frac{\rho_0}{3}}}\frac{1}{\pi\nu}EllipticE(\pi\tau\nu
+\frac{\theta}{2},2)]},
\end{eqnarray}
where EllipticE is an elliptic integral of second kind.

The EoS of the model is
\begin{equation}
\omega=-\,{\frac {3\, \left( \cos \left( 2\,\pi \,\nu\,\tau+\theta
\right)
 \right) ^{2}\sqrt {\varrho _{{0}}}-4\,\sqrt {3}\sin \left( 2\,\pi \,
\nu\,\tau+\theta \right) \pi \,\nu}{3 \left( \cos \left( 2\,\pi
\,\nu\,\tau+ \theta \right)  \right) ^{2}\sqrt {\varrho _{{0}}}}} .
\end{equation}
One can find a new set of coordinates ${x,y,z}$ such as (40)-(42),
and show that there are some knot like figures in a configuration
space of these coordinates.
\section{Conclusion}
Going to some mathematical structures, the knot theory, we studied
the relations between some oscillatory solutions of the FRW
equations and the geometrical picture of some types of the knot. We
show that, by assuming the periodic forms for pressure and energy
density as a functions of time, there exists a coordinate set, in
which the time evolutions of the space is knot like. This formal
similarity repeated when we examined other types of the matter
density and the pressure. Our work, exhibited some interesting
features of the rich, hidden, mathematical structures of the
non-linear FRW equations. Also, we discussed some models, described
the existence of a cyclic universe. We obtained the exact solutions
for the scale factor, the EoS parameter $\omega$ is all models. Also  we have considered some examples knot universes for the Bianchi - I spacetime. Finally it is interesting to extend the results of this paper to the F(R) and F(G)  gravity theories (see e.g. the ref. \cite{sergey1}-\cite{sergey2}) as well as F(T) gravity \cite{T1}-\cite{?}.

  \end{document}